\documentclass[runningheads]{llncs}
\usepackage[T1]{fontenc}
\usepackage{amsmath,amssymb,amsfonts,bbm,xcolor}
\usepackage{algorithm}
\usepackage{algorithmic}
\usepackage{subcaption}
\usepackage{graphicx}
\usepackage{svg}
\usepackage{multirow}
\begin{document}
\title{When There Is No Decoder: Removing Watermarks from Stable Diffusion Models in a No-box Setting}
\titlerunning{When There Is No Decoder}
%
\author{Anonymous Submission}
\author{Xiaodong Wu\inst{1} \and
Tianyi Tang\inst{1} \and
Xiangman Li\inst{1} \and
Jianbing Ni\inst{1}   \and
Yong Yu\inst{2}} 

\authorrunning{X. Wu et al.}
%
\institute{Queen's University at Kingston, Kingston, Ontario, Canada, K7L 3N6
\email{\{xiaodong.wu, 23ch, xiangman.li, jianbing.ni\}@queensu.ca}
\and
School of Computer Science, Shaanxi Normal University, Xi'an 710119, China \email{yuyong@snnu.edu.cn}}  
%
\maketitle              
\begin{abstract}
Watermarking has emerged as a promising solution to counter harmful or deceptive AI-generated content by embedding hidden identifiers that trace content origins. However, the robustness of current watermarking techniques is still largely unexplored, raising critical questions about their effectiveness against adversarial attacks. To address this gap, we examine the robustness of model-specific watermarking, where watermark embedding is integrated with text-to-image generation in models like latent diffusion models. We introduce three attack strategies: edge prediction-based, box blurring, and fine-tuning-based attacks in a no-box setting, where an attacker does not require access to the ground-truth watermark decoder. Our findings reveal that while model-specific watermarking is resilient against basic evasion attempts, such as edge prediction, it is notably vulnerable to blurring and fine-tuning-based attacks. Our best-performing attack achieves a reduction in watermark detection accuracy to approximately 47.92\%. Additionally, we perform an ablation study on factors like message length, kernel size and decoder depth, identifying critical parameters influencing the fine-tuning attack’s success. Finally, we assess several advanced watermarking defenses, finding that even the most robust methods, such as multi-label smoothing, result in watermark extraction accuracy that falls below an acceptable level when subjected to our no-box attacks.
\end{abstract}
\section{Introduction}
\label{sec:intro}


Although generative AI holds the potential to boost efficiency and tackle capacity limits, it also raises critical questions about human creativity, originality, and copyright \cite{guzik2023originality,wu2024unveiling}. Distinguishing AI-generated content (AIGC) from human-generated content is increasingly difficult, complicating copyright issues, especially as misuse of models like Stable Diffusion \cite{rombach2021highresolution} on social media leads to fake images and face-swapping of celebrities. To counter such misuse, watermarking techniques \cite{zhao2023provable,xing2024assessing} enable the traceability of AIGC by embedding identifiable marks. These watermarks, created by a pre-trained encoder and detected by a corresponding decoder, verify an image's origin by embedding and later recognizing predefined watermarks. Most watermarking approaches \cite{jiang2024watermark} rely on an autoencoder structure with encoder and decoder models and can be divided into model-specific and data-specific techniques. Model-specific methods \cite{fernandez2023stable,xiong2023flexible} integrate the encoder directly into the generative model, producing watermarked images during inference without additional processing. Data-specific methods \cite{zhu2018hidden,tancik2020stegastamp} apply the watermark after generation, keeping the encoder separate from the generative model.

Despite their potential, watermarking techniques are vulnerable to adversarial attacks. Simple evasion methods, such as cropping, rotation, or brightness adjustment \cite{an2024benchmarking}, reduce watermark visibility but often introduce noticeable image alterations, limiting their real-world applicability. More sophisticated attacks, often performed in a white-box setting, leverage full access to the watermark decoder. For example, methods proposed by Jiang et al. \cite{jiang2023evading} and Lukas et al. \cite{lukas2023leveraging} use gradient information to evade the detection. However, white-box attacks are rarely feasible in practice due to strict access restrictions on proprietary watermarking decoders.
To address this, black-box attacks are proposed with the attempt to circumvent these access restrictions; however, they come with their own set of limitations. For instance, Jiang et al.’s HopSkipJump \cite{chen2020hopskipjumpattack} method requires frequent queries to the decoder, which undermines stealth by creating identifiable patterns of activity that could be flagged as suspicious. Kassis et al.’s spectral optimization method \cite{kassis2024unmarker} is another black-box approach that can effectively remove watermarks; however, it often causes slight visual degradation, reducing the overall quality and authenticity of the image. In conclusion, existing approaches have limitations due to their reliance on varying degrees of decoder access. None of these methods address a no-box setting, where watermark removal could be achieved without any dependence on decoder access, highlighting a gap in practical and minimally invasive solutions for bypassing watermarks in generative models.

This paper examines the robustness of model-specific text-to-image watermarking techniques in a no-box setting, where attackers have no access to the watermark decoder. We propose three attacks aimed at removing watermarks while preserving image quality, assessing watermark resilience under these real-world constraints. First, we introduce an edge-based attack that manipulates image edges to disrupt watermarks, which cannot remove the watermark without damaging the quality of the target image. Next, we develop a blurring technique that applies box blurring to remove the watermark, followed by deblurring to restore image quality, achieving a watermark detection accuracy of around 0.5 while maintaining visual fidelity. Finally, we explore a fine-tuning attack, leveraging the widespread practice of fine-tuning generative models for custom applications (e.g., specialized fields or unique styles). This approach modifies model weights using a surrogate watermark decoder, effectively erasing watermarks with no noticeable loss in image quality. 
Unlike previous research, which often depends on a white-box setting (with full decoder access) or a black-box setting (with query-based decoder access), our approach introduces a novel framework that completely removes reliance on the decoder. Through this no-box analysis, we evaluate the effectiveness of each attack method and provide insights into the inherent vulnerabilities of model-specific watermarking, contributing a new perspective to watermark resilience research in generative AI systems.
Our contributions are as follows: 
\begin{itemize} 
\item We propose and implement three distinct attack strategies: edge-based, box blurring, and fine-tuning, to systematically assess the robustness of generative watermarking techniques in a no-box setting. 
\item Through extensive experiments, we show that while current watermarking methods withstand simple edge-based attacks, they exhibit significant vulnerabilities to box blurring and fine-tuning, resulting in a marked drop in watermark bit accuracy to 47.92\%. 
\item We evaluate three recently proposed defense methods aimed at enhancing watermark robustness, finding that although they mitigate some attack impact, watermark extraction accuracy still falls below an acceptable threshold, underscoring the effectiveness of our no-box attacks.
\end{itemize}

\section{Related Works}
\label{sec:related}

In this section, we review the existing watermark techniques and attacks on the generative text-to-image models. 

\subsection{Watermarking Techniques on Text-to-image Models}
To protect copyright for images generated from text-to-image models, invisible watermarks can be embedded using autoencoder networks, where an encoder embeds a watermark and a decoder later extracts it. Watermarking methods are either model-specific or data-specific, depending on whether the encoder is integrated with the generative model. 
In model-specific methods \cite{liu2023watermarking,peng2023intellectual}, the encoder is needed only during training, and the watermarks are embedded directly into the images at inference. 
A pioneering approach using this method was proposed by Fernández et al. \cite{fernandez2023stable}, who integrated the encoder into the state-of-the-art text-to-image model, Stable Diffusion.
Inspired by this work, Xiong et al. \cite{xiong2023flexible} designed an improved network capable of embedding a flexible, user-assigned message into each generated image, rather than generating images with the same watermark once the diffusion model is fine-tuned. This improvement was achieved by appending additional modules to the original diffusion model, enabling the embedding of messages during the generation process. 

In data-specific methods \cite{ma2023generative,liu2024countering}, the encoder is not integrated into the generative model itself. Instead, an original image is first generated by a clean generator and then watermarked by a pre-trained encoder \cite{koh2017understanding}. Zhu et al. \cite{zhu2018hidden} pioneered this approach with their HiDDeN framework, which watermarks images using a convolutional neural network (CNN) to construct the encoder and decoder. In their method, original images and messages are input into the encoder, followed by the addition of a special noise layer to simulate real-world image distortions, such as Gaussian noise, cropping, and JPEG compression. The decoder is then trained together with the encoder by optimizing an image reconstruction loss. Despite its effectiveness, HiDDeN's generalization ability remains limited.
To address this limitation, Zeng et al. \cite{zeng2023securing} proposed a method to inject a universal adversarial signature into generated images by training a universal signature injector against a binary signature classifier adversarially. 

\subsection{Malicious Attacks on Image Watermarking}
To target and corrupt the latent watermarks injected into images, numerous attacking methods have been proposed. These methods can be broadly categorized into post-processing methods and learning-based methods.
In post-processing methods, images are manipulated directly in the pixel space. For instance, as demonstrated in \cite{an2024benchmarking}, attackers can perform actions such as rotation, resized cropping, erasing parts of the images, or altering the brightness and contrast. They can also apply Gaussian blur or noise and perform JPEG compression on the target image. The rationale is that since the watermark is embedded in the pixel space, degrading the image quality through these manipulations will also corrupt the embedded watermark.
Additionally, since most current generative models, such as those based on diffusion models, generate images in the latent space using U-Net and VAE models, attacks can also be designed to modify the latent features. For example, \cite{zhao2023invisible} proposed adding noise to the latent representations used to generate images. This approach aims to invalidate the hidden watermark while preserving the quality of the generated images.

In learning-based methods, attackers build additional modules to evade watermark decoders. For instance, Jiang et al. \cite{jiang2023evading} proposed WEvade, which can successfully attack model-specific watermark techniques with dual-tail decoders in both white-box and black-box settings.
In the white-box setting, the decoder is accessible to the attacker, allowing them to obtain gradients of any input. This enables the construction of evasion attacks. Given a watermarked image, attackers can input it into the decoder and use an assigned fake message to guide modifications to the image.
In the more challenging black-box setting, where the decoder is not accessible to attackers, Jiang et al. first designed a surrogate model-based method. Here, a surrogate decoder is trained to simulate the target decoder. However, due to potential differences between the two decoders, the attack performance might be suboptimal. To address this, they proposed a query-based method using a state-of-the-art hard label query-based adversarial approach called HopSkipJump. This approach evades the target decoder with limited query access.
Building on this work, Hu et al. \cite{hu2024transfer} improved attack performance in the black-box setting by employing transfer attack techniques. Specifically, they built multiple surrogate decoders and manipulated a target watermarked image to evade all these decoders simultaneously. With a sufficiently large number of surrogate decoders, the overlap between these decoders and the target decoder enhances attack performance.
Additionally, Lukas et al. \cite{lukas2023leveraging} addressed the problem in a white-box setting by applying a differentiable surrogate key to facilitate the attack. 

\section{Our Attacks on Watermarks}
In this section, we present three attack methods under a no-box setting, aiming at compromising the security of watermarked images: edge prediction, box blurring, and fine-tuning. We begin by outlining the threat model that underpins these attacks, followed by a formal definition of the watermark disruption problem. Each attack is then described in detail, highlighting its specific approach to removing or obscuring the watermark.

\begin{figure}[tb]
\centering
\includegraphics[width=\textwidth]{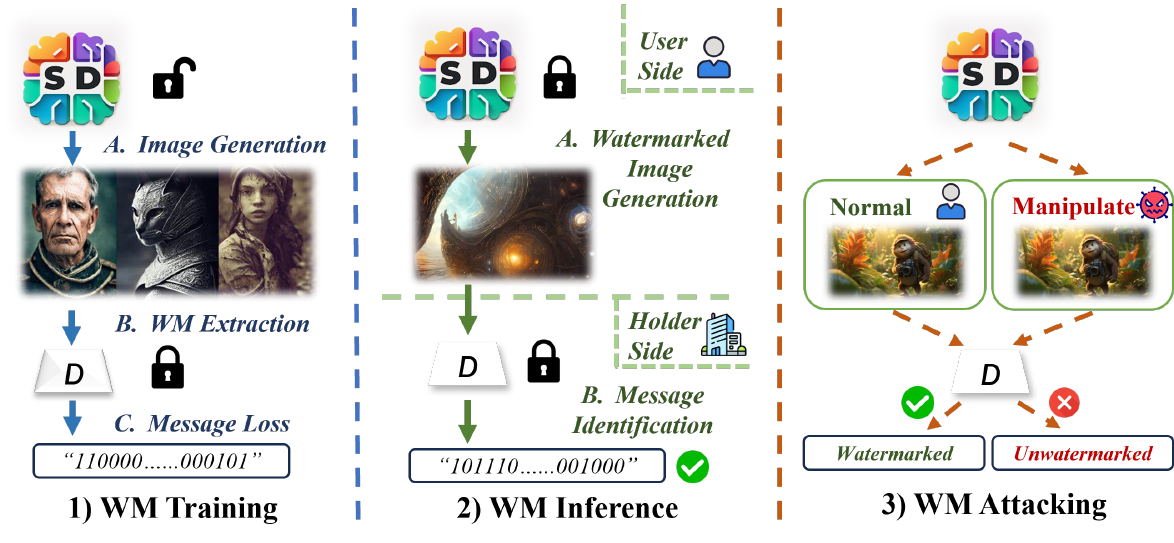}
\caption{Watermark (WM) generation and attack scenarios}
\label{fig:scenario}
\end{figure}

\subsection{Threat Model}
In our attack scenario, as shown in Fig. \ref{fig:scenario}, text-to-image generators provided by model vendors are equipped with watermarking techniques to ensure content authenticity. We focus specifically on model-specific watermarking methods, where watermark encoders are integrated into the generators, as they offer a subtle way to authenticate images and are more straightforward for users to implement. In a learning-as-a-service (LAAS) scenario, these generators create watermarked images by processing text prompts, which describe the image content, along with a binary watermark string that is embedded within the generated images to confirm authenticity. The embedded watermarks can be extracted as multi-bit binary messages to verify the image’s legitimacy.

Regarding adversaries, they are assumed to have a foundational understanding of the watermarked text-to-image generator’s capabilities, including awareness that the generator embeds watermarks using proprietary techniques. These adversaries can fine-tune the generator with their own surrogate watermark decoder. They possess access to a fine-tuning dataset, as well as the generator’s weights, gradients, and necessary computational resources.
However, attackers are restricted to operating in a no-box setting. Specifically, they have no access to the groundtruth decoder employed by the generator’s owner (referred to as the target decoder). These components are treated as fully opaque, meaning attackers lack any knowledge of their internal structure, such as layer count, configuration details, or specific design features, and cannot interact with them directly in any way. Furthermore, the exact bit length and content of the embedded watermark messages remain confidential, further obscuring the watermarking mechanism from the adversary.
This no-box setting presents significant challenges for watermark removal, as attackers must rely solely on manipulating the generator or external image manipulation without access to or feedback from the proprietary decoder, making the removal process indirect and substantially more difficult.

\subsection{Problem Formulation}
Consider an input prompt $s = \{w_1, w_2, \cdots, w_n\}$, which consists of $n$ words, and let $G(\cdot)$ represent the image generation model. The image generation process can be formalized as: 
\begin{equation}
    I = G(s, r, st; \theta_G),
\end{equation}
where $I$ is the generated image, $r$ is a random seed controlling stochastic variations in the output, $st$ denotes the number of denoising steps, and $\theta_G$ represents the parameters of the generator $G(\cdot)$.

Given that the generator has successfully embedded a watermark, the generated image $I$ must meet the following requirement:
\begin{align}
    Sim(m, \bar{m}) > \lambda, \\
    \bar{m} = Dec(I; \theta_D). 
\end{align}
Here, $Dec(\cdot)$ is the watermark decoder that extracts the embedded message $\bar{m}$ from the generated image $I$. The parameters of the decoder are represented by $\theta_D$, and $m$ is the intended message that the generator should embed within $I$. For the image to be verified as produced by this generator $G(\cdot)$, the similarity $Sim(m, \bar{m})$ between the decoded message $\bar{m}$ and the original message $m$ must exceed a threshold $\lambda$.

The attacker's goal is to prevent the generated image $I$ from being recognized as AIGC. To accomplish this, the attacker is allowed to modify both the image $I$ and the generator $G(\cdot)$ as long as the new output image $\hat{I}$ appears visually unchanged from $I$. Drawing on the insights from \cite{jiang2023evading}, instead of reducing the similarity $Sim(m, \hat{m})$ to zero, a more effective approach is to adjust it to approximate $0.5$. This strategy reduces the likelihood of detection, even in cases of double-sided verification. The attacker's objective function can therefore be defined as follows:
\begin{equation}
\begin{split}
    |Si&m(m, \hat{m}) - 0.5| < \hat{\lambda}, \\
    &\hat{m} = Dec(\hat{I} ; \theta_D), \\
    &\hat{I} = G(s, r, st ; \hat{\theta}_G),
\end{split}
\end{equation}
where $\hat{I}$ and $\hat{\theta}_G$ represent the altered image and modified generator weights, respectively, and $\hat{\lambda}$ is a predefined threshold that constrains the similarity between the new message $\hat{m}$ and the original message $m$.

\begin{figure}[tb]
\centering
\includegraphics[width=0.8\textwidth]{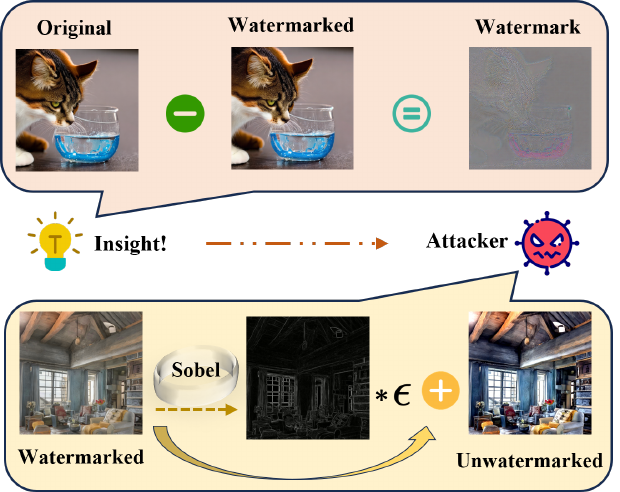}
\caption{An example of the comparison between an original image and its corresponding watermarked image, as well as an illustration of our edge prediction attack.}
\label{fig:edgea}
\end{figure}

\begin{figure}[tb]
\centering
\includegraphics[width=0.9\textwidth]{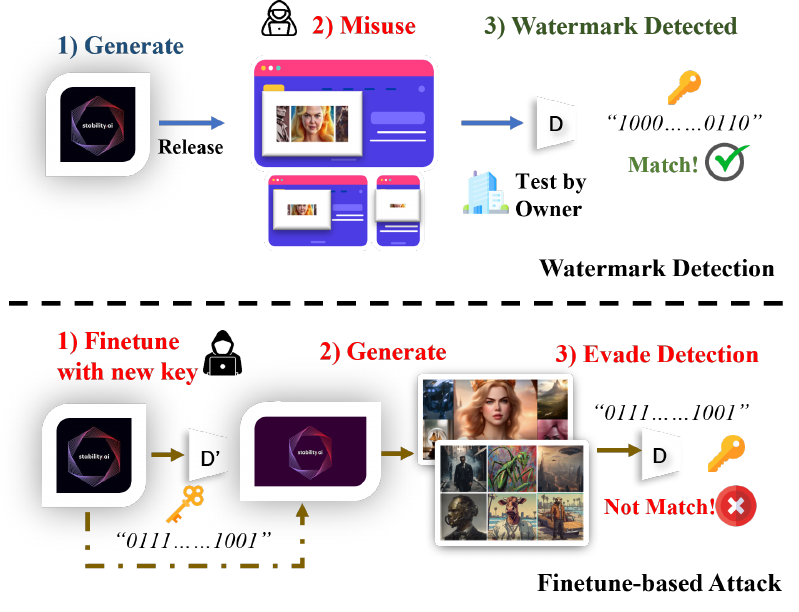}
\caption{Fine-tune-based attack against watermark detection.}
\label{fig:attack_stru}
\end{figure}

\subsection{Edge Prediction-based Attacks}
As illustrated in Fig. \ref{fig:edgea}, the visual difference between watermarked and unwatermarked images is almost imperceptible, and there are subtle alterations near the edges. Based on this observation, we propose an edge-based prediction method that specifically targets these areas. The method first identifies the edges in the watermarked image and then introduces noise into those regions. Formally, given a watermarked image $I$, the attack proceeds as follows:
\begin{equation}
\begin{split}
    &e = Edge(I), \\
    &\mu = \epsilon \cdot \mathbbm{1}(e), \quad \epsilon \sim \mathcal{N}(0, 1), \\
    &\hat{I} = I + \mu,
\end{split}
\end{equation}
where $Edge(\cdot)$ is an edge prediction algorithm, e.g., Sobel \cite{gao2010improved} and $\mathbbm{1}(\cdot)$ is an indicator function that outputs a binary mask corresponding to the detected edges. The noise $\epsilon$, drawn from a standard normal distribution $\mathcal{N}(0, 1)$, is applied to the edge areas to perturb the watermark while minimizing noticeable artifacts in the rest of the image.

\subsection{Blurring Attacks}
Building on the observation that watermarks are often embedded in the edge areas of images, we propose a second method for removing the watermark by modifying these regions. To achieve this, we apply the box blurring technique, a simple yet effective method in image processing for smoothing images and reducing fine scale details. Box blurring works by averaging the pixel values in the local neighborhood of a target pixel. In box blurring, all pixels within the neighborhood have equal weight, resulting in a uniform blurring effect across the region. The box blur kernel for an image is a square matrix of size $n \times n$, where $n$ is a positive integer that determines the extent of the blur (larger $n$ results in more blur). Each element $h(i, j)$ of the box blur kernel $\mathcal{H}$ is defined as:

\begin{equation}
h(i,j)=\frac{1}{n^2}, \quad \text{for } i,j = 0, 1, \cdots, n-1.
\end{equation}

When applying the box blur to an image $I(x, y)$, the blurred image $B(x,y)$ is calculated through a convolution operation with a normalized average over an $n \times n$ neighborhood:

\begin{equation}
B(x,y) = \sum_{i = 0}^{n-1} \sum_{j = 0}^{n-1} I\bigl(x + i - \lfloor n/2 \rfloor, y + j - \lfloor n/2 \rfloor\bigr) \times h(i,j).
\end{equation}

Since $h(i,j) = \frac{1}{n^2}$ for all $(i,j)$ within the kernel, the formula simplifies to:

\begin{equation}
B(x,y) = \frac{1}{n^2} \sum_{i = 0}^{n-1} \sum_{j = 0}^{n-1} I\bigl(x + i - \lfloor n/2 \rfloor, y + j - \lfloor n/2 \rfloor\bigr).
\end{equation}

After blurring, we apply FFTformer \cite{kong2023efficient}, a recent deblurring technique, to recover the image’s original details. FFTformer uses a Discriminative Frequency Domain-based Feedforward Network (DFFN) with a gated mechanism inspired by the Joint Photographic Experts Group (JPEG) compression to selectively retain essential low- and high-frequency features, effectively restoring image clarity and fine details.

\subsection{Fine-tune-based Attacks}
The third approach is to manipulate the generator itself to produce unwatermarked images that retain original features. In model-specific watermarking, pre-trained generators are fine-tuned with a watermark decoder to embed specific messages. Successful watermarking occurs when the decoder accurately retrieves this embedded message. To bypass this, as shown in Fig. \ref{fig:attack_stru}, we propose fine-tuning the generator with an unrelated message so that the decoder retrieves this new message instead, effectively neutralizing the original watermark. This fine-tuning attack targets pre-watermarked models, not unaltered ones. Specifically, this fine-tune-based attack can be described as follows:
\begin{equation}
\begin{split}
    \min_{\theta_G} ; \left| FDec\bigl(G(s, r, st; \theta_G); \theta_D\bigr) - \hat{m} \right|_2, \
\text{s.t.} \quad \hat{m} \neq m_0,
\end{split}
\end{equation}
where $m_0$ and $\hat{m}$ represent the original assigned watermark message and the adversarial message that the generator is fine-tuned to embed. Here, $FDec(\cdot)$ denotes a fake decoder pre-trained by the attacker to approximate the target's actual watermark decoder, which is typically unknown. The success of this attack depends mainly on two factors: The selection of the new adversarial message $\hat{m}$ and the accuracy of $FDec(\cdot)$ in replicating the original decoder’s behavior. A well-constructed $FDec(\cdot)$ increases the likelihood of embedding $\hat{m}$ effectively and evading the original watermark detection.

\section{Experiments}
In this section, we start by describing the dataset used to pre-train the decoder, as well as to fine-tune the generator. We then outline the metrics used to evaluate the effectiveness of our attack methods. Finally, we analyze the performance of the three proposed attack methods, including results from ablation studies to assess the contribution of each approach and evaluate the attack performance in the presence of defense mechanisms.

\subsection{Experimental Setup}
\subsubsection{Datasets.}
We use the Common Objects in Context (COCO) dataset to pre-train the decoder following the approach in \cite{fernandez2023stable}. The COCO dataset, introduced by Lin et al. \cite{lin2014microsoft}, is extensively used in tasks such as object detection, segmentation, and captioning. It contains over 330,000 images, including 200,000 labeled images across 80 object categories, with detailed annotations for segmentation, object labels, and contextual relationships. For our experiments, we randomly select approximately 500 images from COCO to fine-tune the generator.

\subsubsection{Evaluation Metrics.}
We evaluate our attack performance using bit accuracy (Acc), Inception Score (IS), Fréchet Inception Distance (FID), and Contrastive Language–Image Pre-training embedding similarity (CLIP). \textbf{Bit Accuracy (Acc)} measures the similarity between the decoded message and the ground-truth message, as outlined in \cite{jiang2023evading}. 
Specifically, it is used to measure the similarity between two binary sequences, i.e., output message $\hat{m}$ from the decoder and a ground-truth message $m$. It is calculated as:
\begin{equation}
    acc = 1 - \frac{diff(\hat{m}, m)}{len(m)},
\end{equation}
where $diff(\hat{m}, m)$ is the number of different bits between $\hat{m}$ and $m$. As illustrated in \cite{jiang2023evading}, only if the bit accuracy is approximately $0.5$, the image be considered unwatermarked. Otherwise, it should be identified as a watermarked image. Here, the attacker's goal is to force the bit accuracy of the modified images to be between $0.23$ and $0.77$.
\textbf{Inception Score (IS) \cite{salimans2016improved}} assesses image quality and diversity, using a pre-trained Inception V3 model to measure the entropy of predicted class labels, where a higher score reflects greater realism and diversity in generated images. \textbf{Fréchet Inception Distance (FID) \cite{heusel2017gans}}  compares the distribution of generated images to real images, using feature statistics (mean and covariance) extracted from Inception V3 \cite{szegedy2016rethinking}. Lower FID scores indicate that the generated images more closely resemble real images in quality. Finally, \textbf{Contrastive Language–Image
Pretraining (CLIP) \cite{radford2021learning}} evaluates the semantic consistency between two images by computing the cosine similarity of their embeddings extracted from a pre-trained CLIP model. Higher scores indicate that the two images share more similar high level semantic content.

\subsubsection{Implementation Details.}
We build the generator by using the Stable Diffusion model \cite{rombach2022high} and the watermark decoder with HiDDen \cite{zhu2018hidden}. Specifically, stable diffusion models are applied with the Hugging Face APIs to generate images with pre-trained weights. Here, the settings of the image generation, like the number of inference step and random seed, can be set up manually. We apply Adam as the optimizer with a learning rate of $0.02$ to pre-train the watermark decoder and $0.0005$ to fine-tune the generator. The implementation of the proposed attacks is conducted on PyTorch over RTX A6000 platform.
We construct the paired dataset used in section 3.3 by inputting different prompts into the stable diffusion model. To be specific, we use $10$ prompts, $10$ random seeds, and $10$ inference steps on both unwatermarked and watermarked models, respectively, resulting in $1,000$ pairs of unwatermarked and watermarked images. For each pair, since two images are generated with the same prompt, random seed, and inference step, they are visually similar. However, the watermark decoder has the ability to classify them by decoding different message from them.
 
\subsection{Evaluation Results}

\begin{figure*}[tb]
\centering
\includegraphics[width=\textwidth]{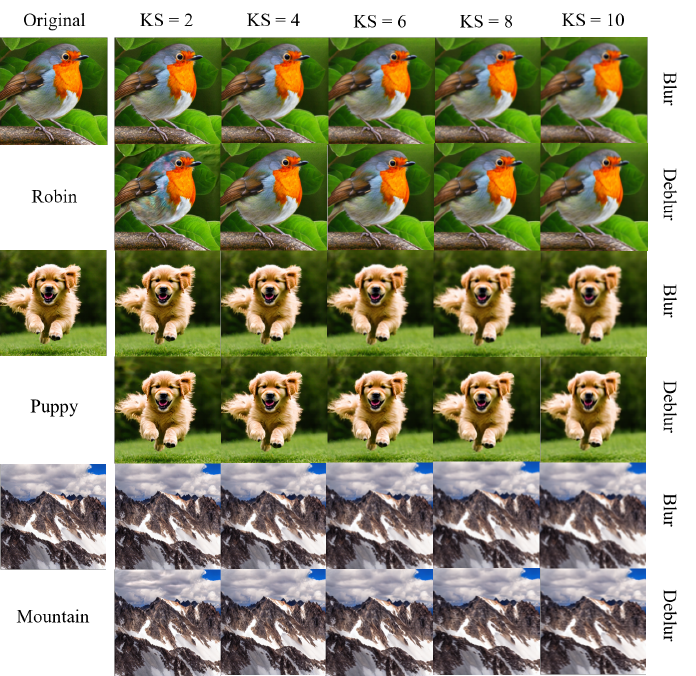}
\caption{{Examples of blurred and deblurred images with varying kernel sizes (`KS' denotes kernel size).}}
\label{fig:birds}
\end{figure*}

\begin{figure}[tb]
\centering

\begin{subfigure}[b]{0.49\textwidth}
    \includegraphics[width=\textwidth]{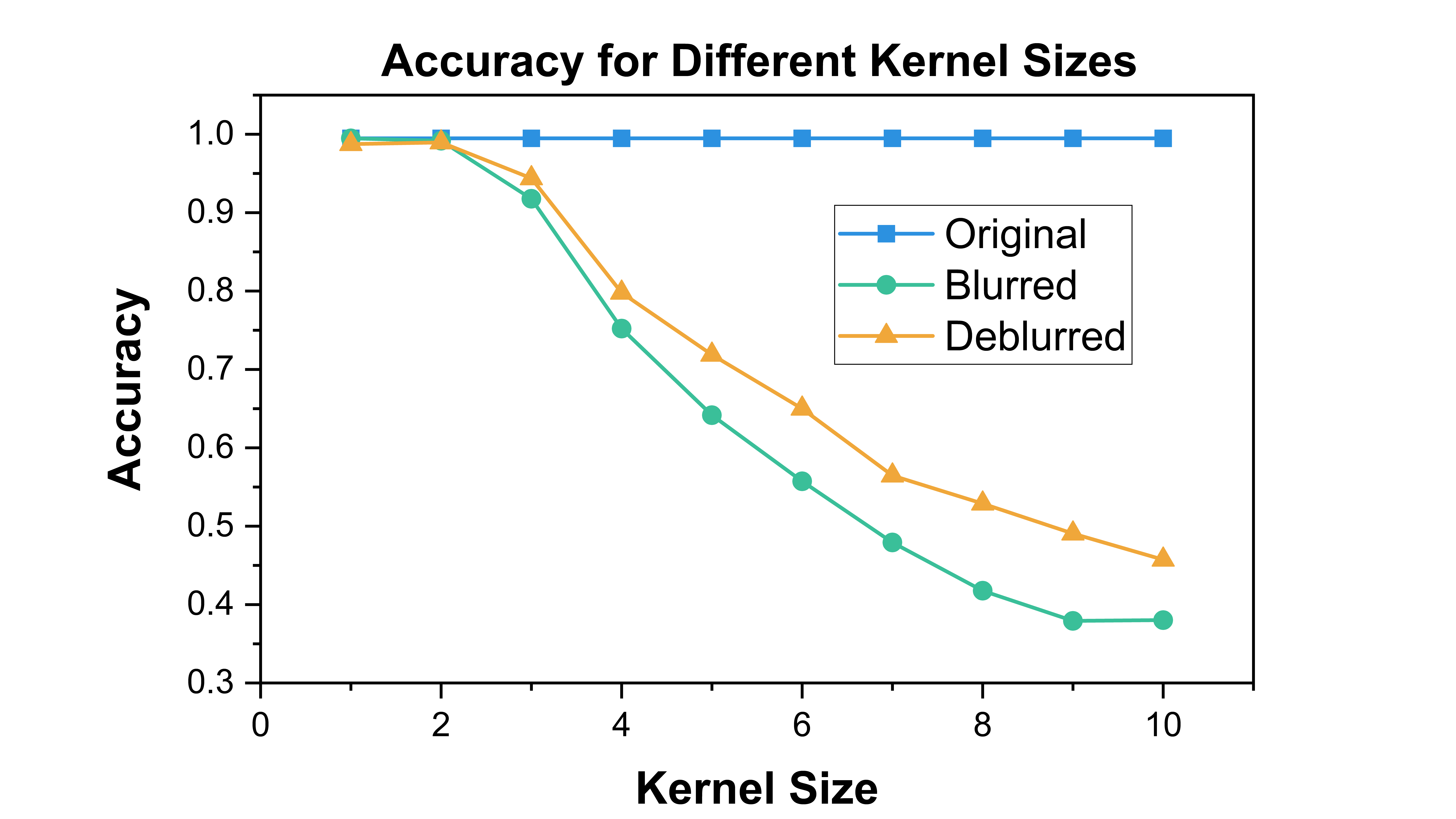}
    \caption{Bit accuracy}
    \label{fig:kernel_acc}
\end{subfigure}
\hfill
\begin{subfigure}[b]{0.49\textwidth}
    \includegraphics[width=\textwidth]{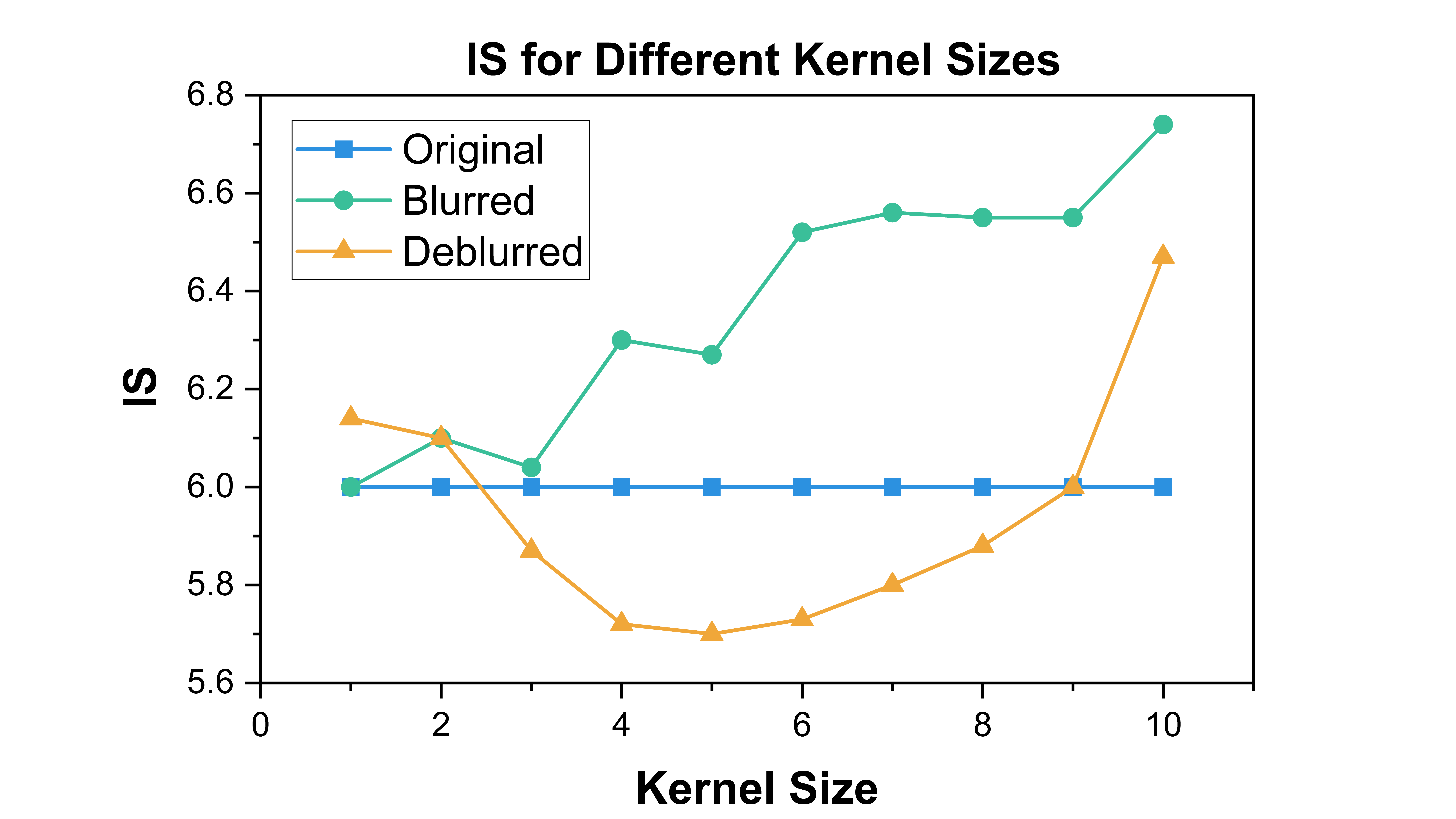}
    \caption{IS}
    \label{fig:kernel_is}
\end{subfigure}
\vspace{0.2cm}
\begin{subfigure}[b]{0.49\textwidth}
    \includegraphics[width=\textwidth]{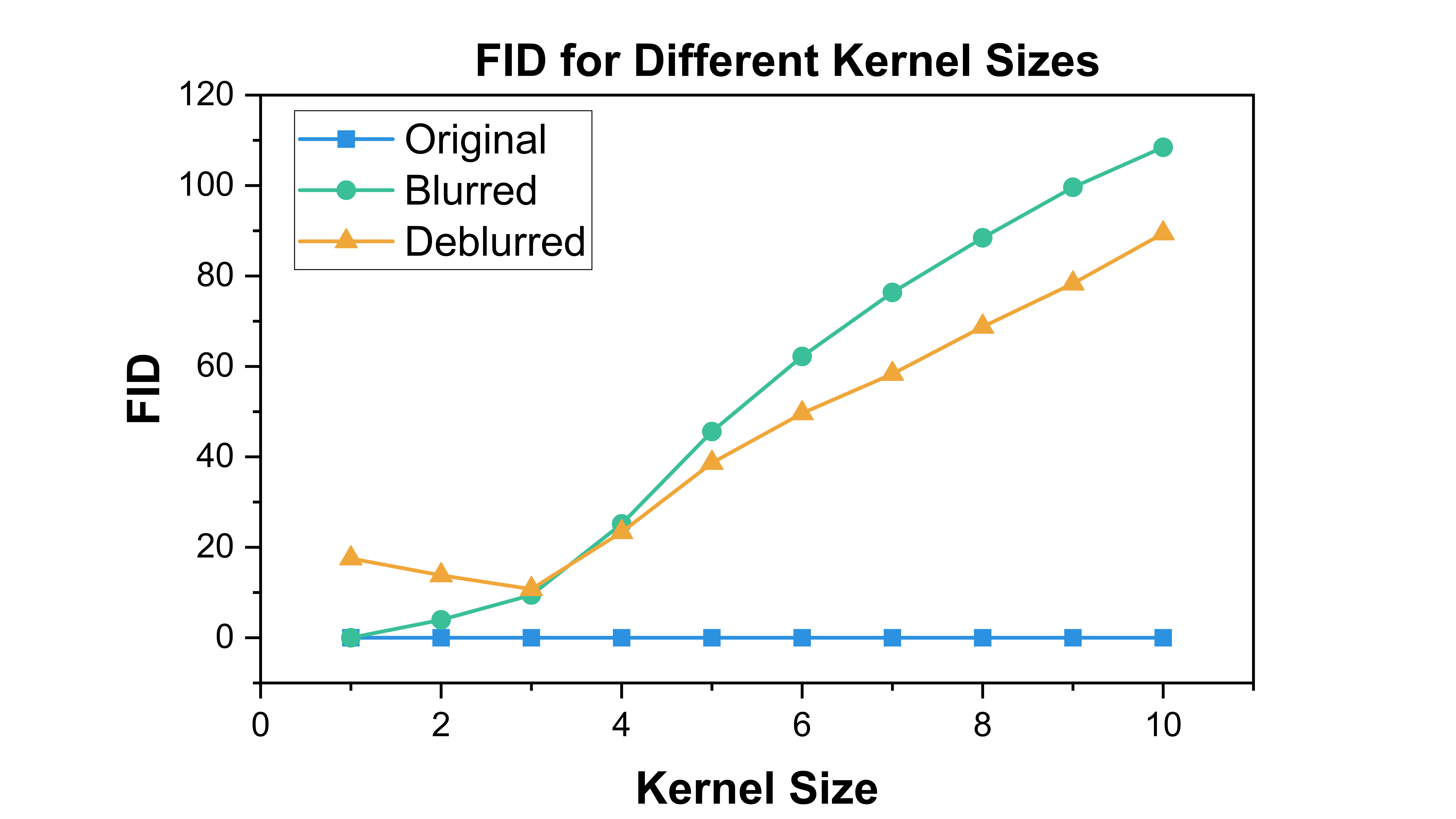}
    \caption{FID}
    \label{fig:kernel_fid}
\end{subfigure}
\hfill
\begin{subfigure}[b]{0.49\textwidth}
    \includegraphics[width=\textwidth]{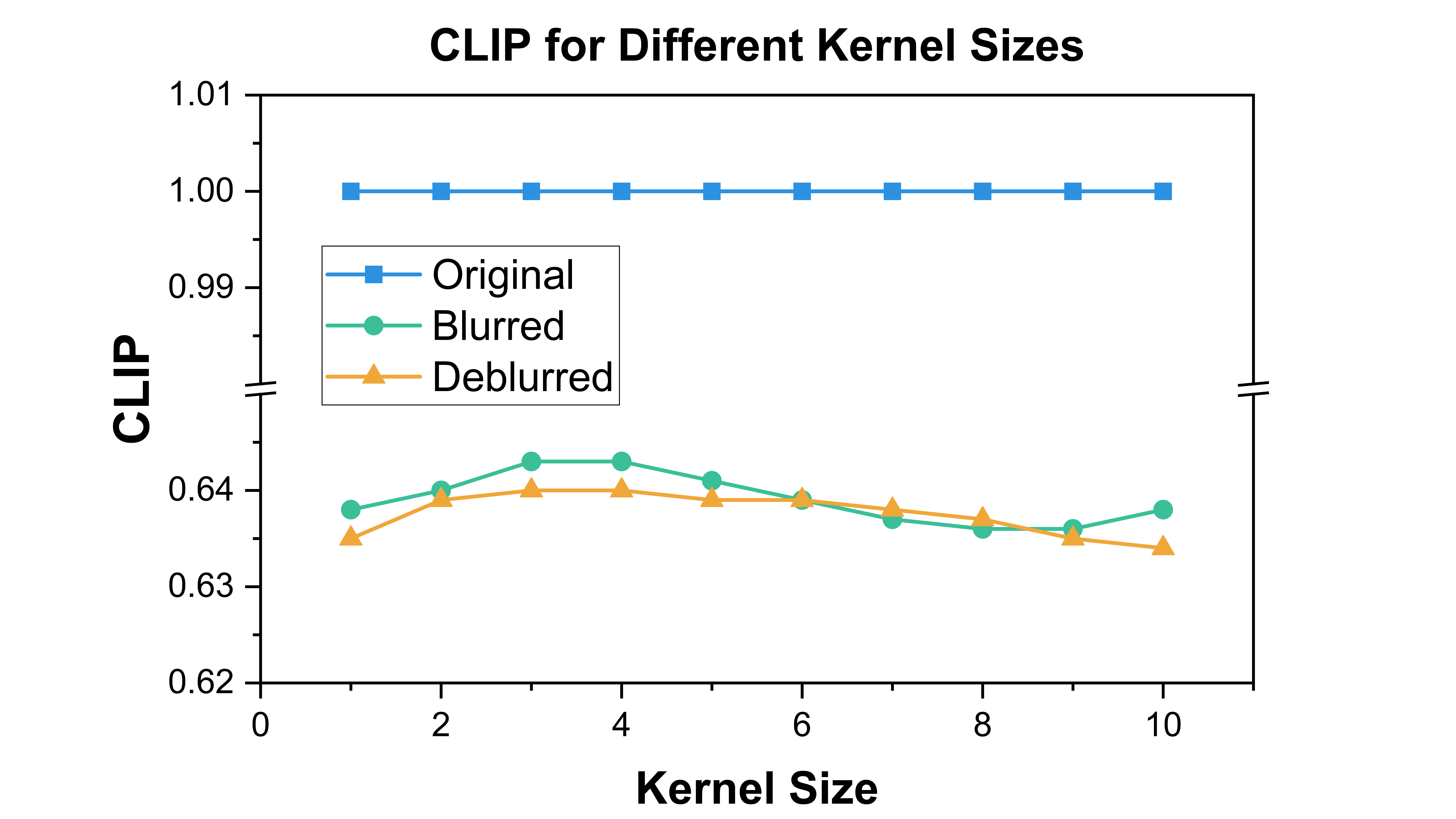}
    \caption{CLIP}
    \label{fig:kernel_clip}
\end{subfigure}

\caption{Bit accuracy, IS, FID, and CLIP embedding similarity for different blurring kernel sizes.}
\label{fig:blur}
\end{figure}

\begin{table}[tb]
    \centering
    \caption{Comparison of bit accuracy, Fid, IS, and CLIP embedding similarity for different blurring methods. The row denoted as "box (k=9)" represents the results of our proposed box blurring attack method when setting the kernel size as 9.}
    \vspace{0.5em}
    \begin{tabular}{c|c|c|c|c|c}
    \hline
     Image Type &Blur Method  &Acc  &Fid  &IS  &CLIP \\\hline
     \multirow{5}{*}{Blurred}
         &8x8  &0.4448  &466.59  &2.31 $\pm$ 0.23&0.530 \\\cline{2-6}
         &16x16  &0.4823  &421.29  &5.53 $\pm$ 1.03 &0.550\\\cline{2-6}
         &32x32  &0.4781  &409.99  &7.36 $\pm$ 0.27&0.566\\\cline{2-6}
         &motion  &0.6031  &84.94  &6.95 $\pm$ 0.18&0.633\\\cline{2-6}
      &gaussian  &0.6271  &54.49  &6.39 $\pm$ 0.21&0.641\\\cline{2-6}
      &\textbf{box (k=9)} &\textbf{0.3792}  &\textbf{99.61}  &\textbf{6.55 $\pm$ 0.33} &\textbf{0.636}\\\hline
     \multirow{5}{*}{Deblurred}
         &8x8  &0.4480  &468.50  &2.33 $\pm$ 0.19&0.529\\\cline{2-6}
         &16x16  &0.4844  &423.54  &5.54 $\pm$ 1.04&0.549\\\cline{2-6}
         &32x32  &0.4823  &410.64  &7.40 $\pm$ 0.25&0.566\\\cline{2-6}
         &motion  &0.8177  &31.23  &6.19 $\pm$ 0.07&0.632\\\cline{2-6}
      &gaussian  &0.6802  &40.53  &5.96 $\pm$ 0.07&0.640\\\cline{2-6}
      &\textbf{box (k=9)} &\textbf{0.4906}  &\textbf{78.33}  &\textbf{6.00 $\pm$ 0.22} &\textbf{0.635}\\\hline
    \end{tabular}
    \label{tab:blur_method}
\end{table}


\subsubsection{Performance of Edge Prediction-based Attacks.}
We evaluate edge prediction attacks based on the observation that watermarked images visually resemble their original counterparts. As shown in Fig. \ref{fig:edgea}, by applying edge detection techniques, such as the Sobel operator \cite{gao2010improved}, we can extract the edges from a watermarked image to generate a grayscale edge map. This map highlights the locations where edges occur in the image, allowing us to selectively introduce noise, such as Gaussian noise, into these regions. Despite the precision of edge detection, our results show that this attack method does not successfully remove the watermark. The manipulated images either retain a high bit accuracy, close to 1.0, indicating that the watermark remains intact, or they become visibly distorted, deviating significantly from the original image. These outcomes reveal that the edge-based injected noise fails to disrupt the watermark effectively without compromising the visual quality of the image.

\subsubsection{Performance of Box Blurring Attacks.}
To assess the effectiveness of our box blurring attacks, we evaluated both blurred and deblurred images using quantitative and qualitative measures. As shown in Figure \ref{fig:birds}, we provide several examples of blurred and deblurred images under different kernel sizes. Visually, we observe that as the kernel size increases, both blurred and deblurred images become increasingly unclear, indicating that larger kernel sizes result in more pronounced blurring effects. When the kernel size is relatively small, i.e., less than 6, the output images still closely resemble the originals.
To further evaluate the effectiveness of our attack on watermark detection and image quality, we calculate the average bit accuracy, IS score, FID, and CLIP across $20$ blurred and deblurred images. The results are shown in Figs. \ref{fig:kernel_acc}, \ref{fig:kernel_is}, \ref{fig:kernel_fid}, and \ref{fig:kernel_clip}. The bit accuracy of both blurred and deblurred image sets follows a decreasing trend as the kernel size grows, demonstrating that larger kernels lead to more effective attacks. Notably, the accuracy of deblurred images consistently surpasses that of blurred images, revealing that deblurring not only enhances image quality but also aids in recovering the embedded watermarks. For the IS and FID scores, we observe that while deblurred images exhibit a lower FID, indicating better alignment with real images, their IS score is also lower, suggesting a decline in image quality or diversity. This discrepancy arises because FID measures global similarity, while IS evaluates distinctiveness and clarity. In other words, while deblurring improves the global structure (reflected in the lower FID), it may introduce artifacts that degrade the clarity and distinctiveness of the images, resulting in a lower IS score. For the CLIP embedding similarity, the reduced value indicates a slight semantic loss introduced by the blurring and deblurring processes compared to the original images.

To evaluate the effectiveness of different blurring techniques in disrupting watermark detection, we conduct an ablation study comparing our proposed \textbf{box blurring} method against four commonly used alternatives. Specifically, we include resize-based methods, where images are downsampled to 8×8, 16×16, or 32×32 and then upsampled back to the original resolution, as well as motion blur and Gaussian blur.
As shown in Table \ref{tab:blur_method}, resize-based methods achieve relatively low bit accuracy (e.g., 0.4448 for 8×8), which indicates a more successful watermark removal attack, as a bit accuracy closer to 0.5 suggests the watermark cannot be reliably extracted. However, this performance comes at the expense of severe image quality degradation, reflected by significantly lower Inception Scores (IS) and higher Fréchet Inception Distances (FID).
Conversely, motion and Gaussian blur preserve image quality much better (e.g., FID of 84.94 and 54.49, respectively), but their higher bit accuracy (0.6031 and 0.6271) indicates that the watermark remains more detectable, thus making the attack less effective.

In contrast, our proposed box blurring method achieves a better trade-off between attack success and image fidelity. It records the lowest bit accuracy (0.3792 in the blurred case), indicating the most effective disruption of watermark detection, while still maintaining competitive image quality (IS of 6.55 and FID of 99.61). This favorable balance persists after deblurring, where our method continues to show superior performance with a bit accuracy of 0.4906 and strong visual quality. Additionally, CLIP embedding similarity confirms that our method preserves semantic content on par with motion and Gaussian blur.
Overall, these results demonstrate that box blurring offers a compelling balance between weakening watermark detection and maintaining visual and semantic integrity, making it a highly effective attack strategy.

\subsubsection{Performance of Fine-tune-based Attacks.}

The fine-tune-based attack corrupts the watermark by fine-tuning the generator again to embed a different watermark in the generated images, thereby preventing the target decoder from correctly identifying the original message. Since the attacker does not have access to the original embedded message, we begin by analyzing how varying the length of the new message affects the success of the attack. Starting with a default message length of 48 bits, we test lengths of 32, 36, 40, 44, 48, 52, 56, 60, and 64 bits, covering cases where the attack message is shorter, equal to, or longer than the default. For each message length, we generate 10 random attack messages and calculate the average bit accuracy. As shown in Table \ref{tab:bit}, shorter attack messages, particularly those of 32 bits, achieve the lowest bit accuracy (64.79\%), while messages closer to the default length of 48 bits yield moderate accuracy (67.92\%). For longer messages (52–64 bits), bit accuracy increases, suggesting a less effective attack, despite improved image quality (lower FID scores). This indicates that while longer messages may enhance image quality, they do not significantly improve the attack's ability to bypass watermark detection.

We next evaluate the impact of the watermark decoder's depth on the success of the attack, considering that the exact architecture of the target decoder is unknown to the attackers. The groundtruth target decoder used in the system has a depth of $8$. To explore how variations in depth influence the attack, we test surrogate decoders with smaller and larger depths and report the results in Table \ref{tab:depth}. The findings indicate that the attack is only successful when the attacker's decoder closely matches the depth of the target decoder. Both underfitting (using a shallower decoder) and overfitting (using a deeper decoder) fail to provide any meaningful advantage in evading the watermark. This suggests that the attack's success heavily relies on the attacker's ability to approximate the correct model capacity of the watermark decoder. Matching the complexity of the target decoder is critical, as deviations whether too simple or too complex, do not enhance the attack's performance.

\begin{table*}[tb]
    \centering
    \caption{Comparison of Bit Accuracy, IS, FID, and CLIP embedding similarity across different fine-tuned message lengths. Class: Multi-Class Smoothing, Label: Multi-Label Smoothing, Regression: Regression Smoothing. The bit length of the ground truth message is equal to 48.}
    \vspace{0.5em}
    \begin{tabular}{c|c|c|c|c|c|c|c}
    \hline
         \#Bit &Acc &IS &FID &CLIP &Class &Label &Regression\\\hline 
         32	&0.6479	&4.91 $\pm$ 0.08 &41.81 &0.618 &0.6750 &0.6708 &0.6760\\\hline
         36	&0.8542	&4.95 $\pm$ 0.04 &38.48 &0.620 &0.9167 &0.9229 &0.9187\\\hline
         40	&0.6521	&4.94 $\pm$ 0.05 &56.85 &0.611 &0.6531 &0.6646 &0.6531\\\hline
         44	&0.7083	&4.82 $\pm$ 0.17 &46.69 &0.622 &0.7563 &0.7625 &0.7573\\\hline
         \textbf{48}	&\textbf{0.6792}	&\textbf{4.85 $\pm$ 0.14} &\textbf{42.98} &\textbf{0.620} &\textbf{0.6865} &\textbf{0.6896} &\textbf{0.6865}\\\hline
         52	&0.7667	&4.87 $\pm$ 0.12 &12.65 &0.630 &0.7917 &0.7896 &0.7937\\\hline
         56	&0.7250	&4.88 $\pm$ 0.11 &15.72 &0.626 &0.7260 &0.7292 &0.7219\\\hline
         60	&0.7292	&4.85 $\pm$ 0.14 &15.00 &0.632 &0.7583 &0.7625 &0.7563\\\hline
         64	&0.7438	&4.83 $\pm$ 0.16 &14.27 &0.632 &0.8104 &0.8313 &0.8104\\\hline
    \end{tabular}
    \label{tab:bit}
\end{table*}

\begin{table}[tb]
    \centering
    \caption{Bit accuracy, IS, FID and CLIP embedding similarity under different decoder depths. The depth of the ground-truth decoder is 8.}
    \vspace{0.5em}
    \begin{tabular}{c|c|c|c|c}
    \hline
         Depth &Acc &IS &FID &CLIP\\\hline 
         4	&0.8938	&4.87 $\pm$ 0.12 &13.99 &0.633\\\hline 
         6	&0.7937	&4.89 $\pm$ 0.10 &14.07 &0.631\\\hline 
         \textbf{8}	&\textbf{0.6792}	&\textbf{4.85 $\pm$ 0.14} &\textbf{42.98} &\textbf{0.620}\\\hline 
         10	&0.6375 &4.83 $\pm$ 0.16 &42.92 &0.616	\\\hline 
         12	&0.7369	&4.92 $\pm$ 0.07 &23.77 &0.625\\\hline 
    \end{tabular}
    \label{tab:depth}
\end{table}

\subsubsection{Defenses.}

\begin{figure}[tb]
\centering
\begin{subfigure}[b]{0.47\textwidth}
    \includegraphics[width=\textwidth]{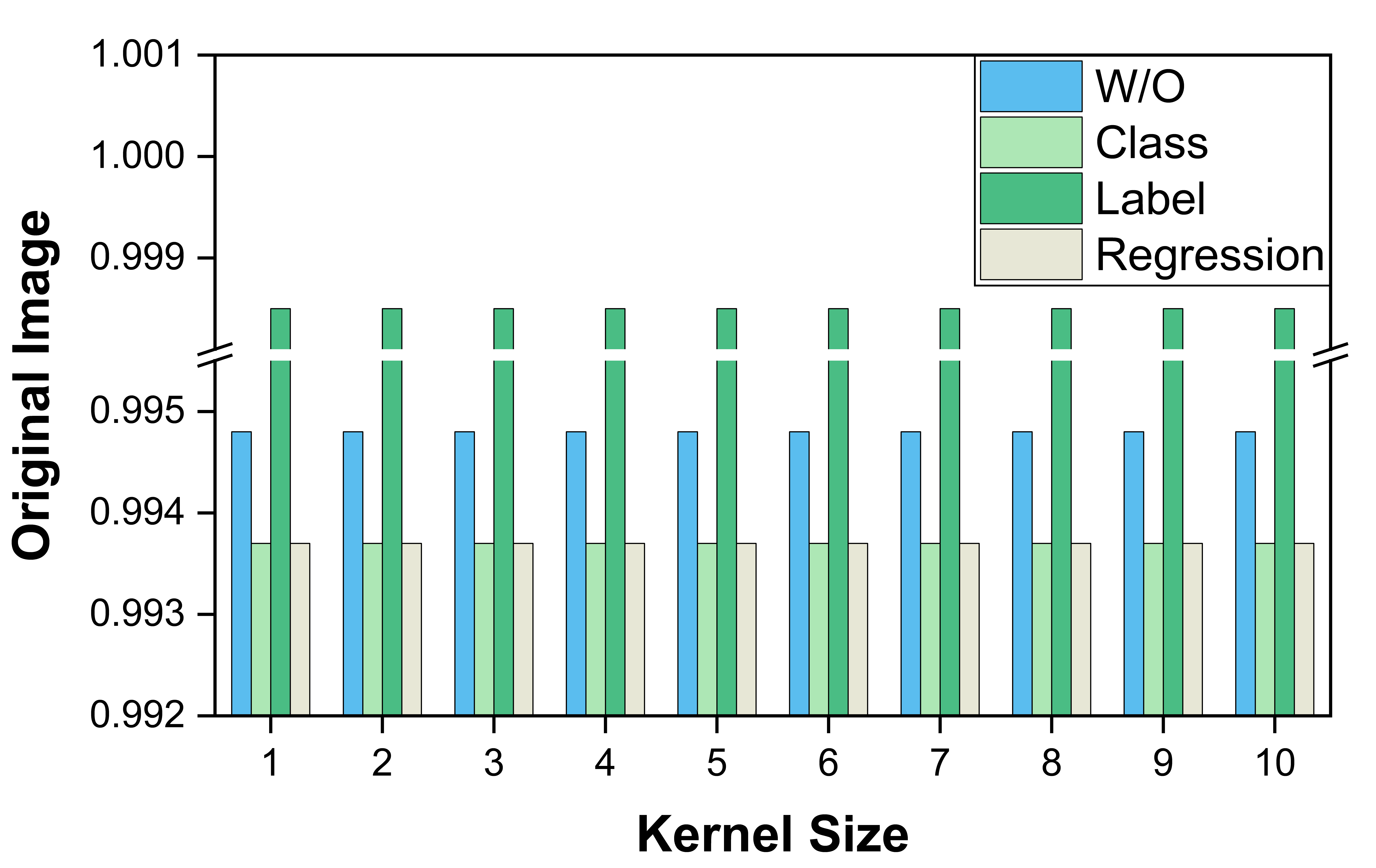}
    \caption{}
    \label{fig:kernel_orig}
\end{subfigure}
\begin{subfigure}[b]{0.47\textwidth}
    \includegraphics[width=\textwidth]{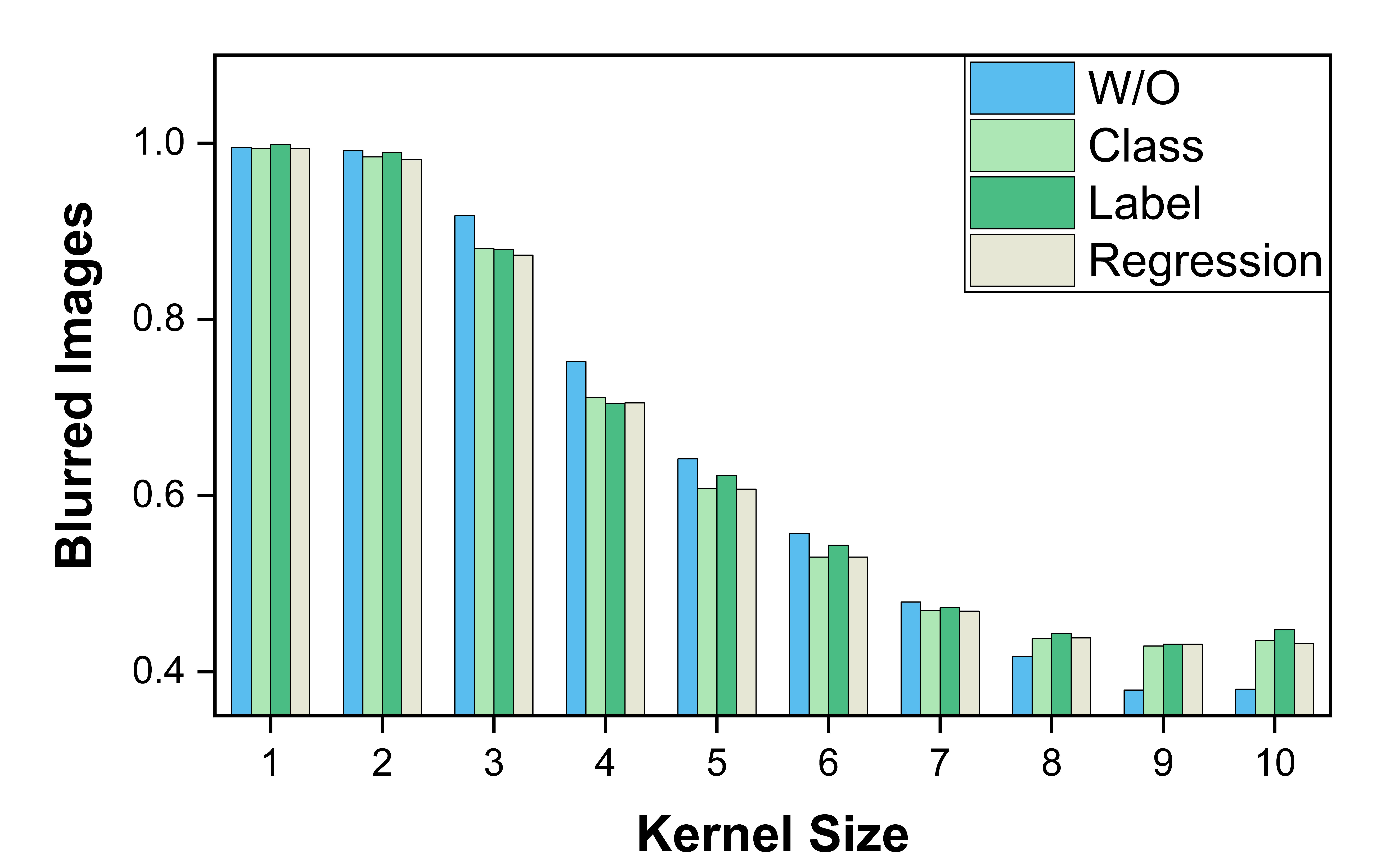}
    \caption{}
    \label{fig:kernel_blur}
\end{subfigure}
\begin{subfigure}[b]{0.47\textwidth}
    \includegraphics[width=\textwidth]{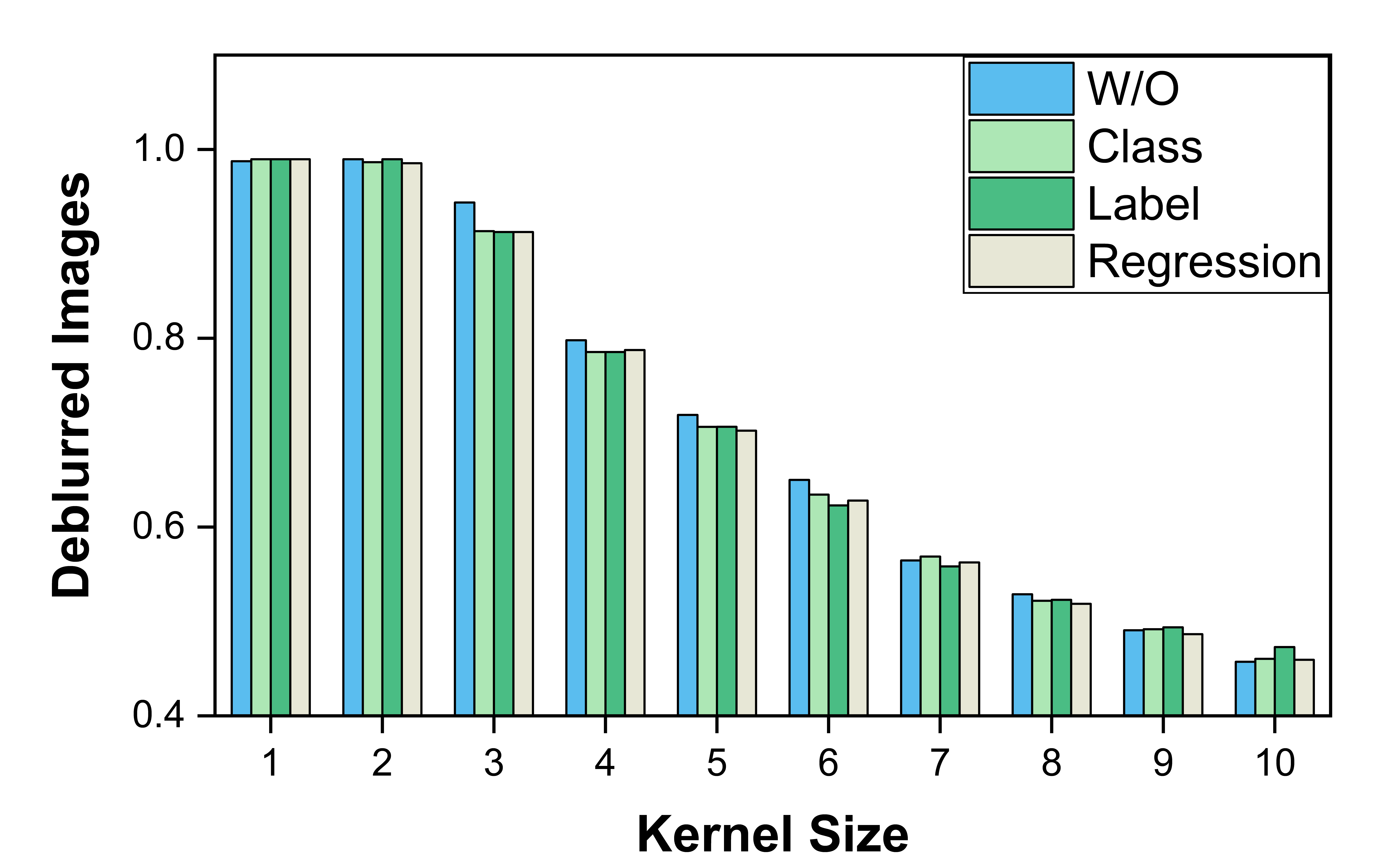}
    \caption{}
    \label{fig:kernel_deblur}
\end{subfigure}
\caption{Comparison among the accuracy of the original images, blurred images, and deblurred images under different blurring kernel sizes.}
\label{fig:defense}
\end{figure}

The vulnerability of model-specific text-to-image watermarking techniques to malicious attacks was demonstrated without the implementation of any defense mechanisms. Now, we consider three novel defense methods recently proposed by Z. Jiang et al. \cite{jiang2024certifiably} for the text-to-image model. Specifically, after generating the watermarked image, we introduce \textit{N} instances of random box noise to create \textit{N} slightly perturbed images. These perturbed images are then fed into the watermark decoder, resulting in \textit{N} watermark messages. Subsequently, three distinct smoothing methods are applied to aggregate these messages and compute the final accuracy for each image: (1) \textit{Multi-Class Smoothing}, which uses a majority vote for each bit; (2) \textit{Multi-Label Smoothing}, which counts the occurrences of each bit as one and sets bits with higher counts to one, and the rest to zero; and (3) \textit{Regression Smoothing}, which calculates bit accuracy for each message and selects the median.

We assess the robustness of the model with these defense methods against box blurring and fine-tune-based attacks. For box blurring attacks, as shown in Figure \ref{fig:defense}, \textit{Multi-Label Smoothing} improves bit accuracy for the original images, while \textit{Multi-Class Smoothing} and \textit{Regression Smoothing} show slight declines. For blurred and deblurred images, the accuracy across all three methods decreases when the kernel size is below eight but improves with larger kernels, suggesting better error correction at higher kernel sizes. In the case of fine-tuning-based attacks (columns 5-7 of Table \ref{tab:bit}), \textit{Multi-Label Smoothing} consistently yields the most significant improvements in bit accuracy across different message lengths, with enhancements up to 0.0875 (from 0.7438 to 0.8313). This indicates that \textit{Multi-Label Smoothing} is particularly effective for improving robustness, likely due to its sensitivity to bit frequency patterns. While \textit{Multi-Class Smoothing} and \textit{Regression Smoothing} also show improvements, \textit{Multi-Label Smoothing} emerges as the most effective defense in countering adversarial changes during fine-tuning scenarios.
However, the accuracy of all three defenses remains below the acceptable threshold in most cases, ranging from 0.27 to 0.73. This suggests that our proposed attacks remain effective even against the latest defense techniques.

\section{Conclusion}
In this paper, we conducted an advanced investigation into the robustness of text-to-image watermarking techniques, focusing on a no-box setting where attackers operate without any access to the groundtruth watermark decoder. We introduced three novel attack strategies: edge-prediction-based, box blurring, and fine-tuning attacks. Our findings revealed that, while existing watermarking methods demonstrate some resilience to basic evasion techniques, they are particularly susceptible to more advanced attacks like box blurring and fine-tuning, even without decoder queries or access.
To counteract these vulnerabilities, we evaluated three state-of-the-art defenses aimed at improving watermark robustness. While Multi-Label Smoothing demonstrated the strongest resilience, its effectiveness remained below an acceptable threshold. 
This highlights the strength of our proposed no-box attacks, which continue to be effective even against state-of-the-art defenses. 
As text-to-image models like Stable Diffusion gain popularity for their openness and adaptability through fine-tuning, our study reveals a critical and growing threat to watermarking security. These findings emphasize the urgent need for more effective defenses to safeguard AIGC from increasingly sophisticated attacks.

%
%
%
 \bibliographystyle{splncs04}
 \bibliography{icics}
\end{document}